# Exchange-Driven Intermixing of Bulk and Topological Surface State by Chiral Excitons in Bi$_2$Se$_3$

Bowen Hou[1], Dan Wang[1], Bradford A. Barker[2], and Diana Y. Qiu[1*]

[1]Department of Mechanical Engineering and Material Sciences,
Yale University, New Haven, CT 06511, USA

[2]Department of Physics, University of California, Merced, CA 95343, USA

Email: diana.qiu@yale.edu

**Abstract**

Topological surface states (TSS) in the prototypical topological insulator (TI) Bi$_2$Se$_3$ are frequently characterized using optical probes, but electron-hole interactions and their effect on surface localization and optical response of the TSS remain unexplored. Here, we use *ab initio* calculations to understand excitonic effects in the bulk and surface of Bi$_2$Se$_3$. We identify multiple series of chiral excitons that exhibit both bulk and TSS character, due to exchange-driven mixing. Our results address fundamental questions about the degree to which electron-hole interactions can relax the topological protection of surface states and dipole selection rules for circularly polarized light in TIs by elucidating the complex intermixture of bulk and surface states excited in optical measurements and their coupling to light.

In the prototypical topological insulator (TI) Bi$_2$Se$_3$, strong spin-orbit coupling (SOC) drives the inversion of bulk bands near the Fermi level, resulting in the appearance of a pair of topological surface states (TSS) with spin-momentum locking[1-13]. These TSS have the potential to be exploited for the generation of topologically-protected spin currents[14], leading to the promise of faster, low-powered spintronics. Due to the high residual conductivity of the bulk[15-20], light-based probes, such as the circular photogalvanic effect, are frequently used to probe the surface states[15,21-26]. In such probes, changes to the photocurrent direction with the helicity of light are interpreted as evidence of photocurrent generation from the TSS. However, the frequency of the probe light is typically 1.5 eV or larger, leading to the excitation of a complex manifold of transitions between numerous bulk and surface states, which can be challenging to disentangle without the help of microscopic theory. For instance, it can be difficult to differentiate photocurrent generated from the TSS near the Fermi level (SS1) and a higher-energy unoccupied topological surface state (SS2), with recent time-resolved angle-resolved photoemission spectroscopy (tr-ARPES) indicating that previous photocurrent measurements likely involve excitations of both states[16]. Furthermore, recent experiments have identified long-lived chiral excitons around 2.4 eV, which are speculated to arise from transitions between a low-energy occupied Rashba surface state (RSS) and SS2[27].

Despite the important role of optical probes in the characterization of TSS, electron-hole interactions leading to the formation of excitons have largely been neglected, with analyses of optical spectra relying on selection rules derived from an independent particle picture. However, recent discovery of long-lived chiral surface excitons[27] raises the question of how excitonic effects can alter the character of the surface excitations. In this work, we utilize the state-of-the-art *ab initio* GW plus Bethe Salpeter equation (GW-BSE) method to investigate the quasiparticle (QP) band structure and optical absorption spectrum including electron-hole interactions for both bulk and quasi-two-dimensional slabs of $Bi_2Se_3$. Our calculations reveal multiple previously unidentified excitons with chiral optical selection rules (chiral excitons) in an energy window from 0-2.82 eV, including bright exciton states with p-like symmetry arising from topologically non-trivial bands[28-30]. Remarkably, these chiral excitons are composed of electrons and holes that combine the character of both bulk and topologically-protected surface states, whose intermixing is driven by the Coulomb exchange interaction. Our calculations address fundamental questions about the degree to which electron-hole interactions change the surface localization of optically-excited TSS, relax the dipole selection rules for circularly polarized light compared to the independent particle picture, and elucidate the complex intermixture of bulk and surface states excited in optical measurements, in good agreement with recent experiments[27,31].

In this Letter, we perform *ab initio* mean-field calculations of bulk, one quintuple layer (QL), 3QL, and 5QL $Bi_2Se_3$ using density functional theory (DFT) in the local density approximation (LDA)[32-34] as implemented in Quantum ESPRESSO [35]. GW and GW-BSE calculations, employing a fully-relativistic spinor formalism[36], as implemented in BERKELEYGW [37-39] are performed on top of the DFT mean field. Computational details can be found in the Supplemental Information (SI). Fig. 1(a) shows the bandstructures of bulk $Bi_2Se_3$.

The top of the valence band at the LDA level has a "camelback" feature, which becomes parabolic at the GW level, consistent with $k \cdot p$ models[1,44] and previous GW calculations[36,42,44-52]. After the many-body electron-electron interactions are included at the GW level, the indirect LDA bandgap of 0.32 eV is renormalized into a direct QP gap of 0.31 eV at the Γ point, which agrees well with previous experimental and theoretical results[42,48]. We note that there is a kink in the one-shot $G_0W_0$ bandstructure along the Γ to Z direction, which becomes smooth when the QP wavefunction is updated self consistently (see SI Fig. S6)[36,46].

Fig. 1 (b-d) shows the bandstructures for 1QL, 3QL, and 5QL slabs of $Bi_2Se_3$. In finite slabs of $Bi_2Se_3$, interaction between surface states at the top and bottom surface open a bandgap at the Γ point in the Dirac cone crossing the Fermi level (SS1) [53]. At the DFT level, the surface state is gapped by 0.53 eV in the 1QL system and becomes nearly metallic in the 3 QL system with a gap 0.017 eV. At the GW level, in 1 QL, the bandgap increases to 1.27 eV, which is 0.95 eV larger than the GW gap of bulk $Bi_2Se_3$. 3QL and 5QL $Bi_2Se_3$ have bandgaps of 0.17 eV and 0.02 eV respectively at the GW level, which agrees well with previous ARPES measurements of 0.15 eV and 0.04 eV[53] and improves on previous GW calculations, where SOC is included perturbatively[44]. Both SS2 and RSS remain ungapped in the 3QL and 5QL slabs at both the DFT and GW levels.

Next, we use the GW-BSE approach to investigate the optical properties of 1QL and 3QL $Bi_2Se_3$. Fig. 2 shows the absorption spectra for 1QL and 3QL $Bi_2Se_3$, with and without electron-hole interactions. Two prominent excitonic peaks, located at 0.8 eV and 1.0 eV, which we label peaks A and B, respectively, appear in the absorption spectrum of 1QL $Bi_2Se_3$. These excitons are strongly bound, with binding energies of 0.46 eV and 0.26 eV, respectively. Peak A consists of contributions from two degenerate bright excitons that are nearly degenerate with a lower energy dark exciton (see SI Fig. S2 for full spectrum), all of which are composed of electron and hole states occupying the doubly degenerate highest valence and lowest conduction bands. We expand each exciton in the electron-hole basis as $|S\rangle = \sum_{vck} A^S_{vck} |vc\bm{k}\rangle$, where $A^S_{vck}$ is the k-space envelope of the wavefunction for the exciton state $S$, composed of electron-hole pairs $|vc\bm{k}\rangle$. For 1QL $Bi_2Se_3$, we plot the phase winding of $A^S_{vck}$ for the two bright excitons in peak A (Fig. 2 (b-c)) along with the phase winding of the dipole matrix element $< ck|\hat{p}_\pm|vk >$ between the highest valence and lowest conduction bands for right ($\hat{p}_+$) and left-hand ($\hat{p}_-$) circularly polarized light (Fig. 2 (d-e)). We find that the lowest energy bright excitons are p-like states with orbital angular momentum *m*=+1 and *m*=-1, while the dipole matrix elements have a winding number of $l_-$=+1($l_+$=-1 ) for left(right)-circularly polarized light. The bright p-like exciton states in conjunction with the non-trivial topology of the underlying independent-particle bands is consistent with the generalized selection rule $m = -l_\mp$ [28-30,54]. The B exciton peak is composed primarily of holes residing in the second-highest valence band and electrons in the lowest conduction band.

Excitonic effects in 3QL $Bi_2Se_3$ are considerably smaller (Fig. 2 (f)) due to its smaller bandgap. The first peak in the absorption spectrum, labelled C, is nearly identical regardless of whether electron-hole interactions are included. We find that the binding energy of the first bright exciton is 10 meV. Since GW significantly increases the bandgap, we perform one self-consistent update of the screened Coulomb interaction W utilizing the GW bandgap of 0.17 eV. After the self-consistent update, the binding energy of peak C increases to 20 meV (see SI Fig. S7). Despite the small binding energy, the optical selection rules of the bound excitons in 3QL $Bi_2Se_3$ follow the same pattern as in 1QL $Bi_2Se_3$: p-like exciton states with orbital angular momentum *m*=-1 and *m*=+1 are bright due to the $l_-$=+1 and $l_+$=-1 phase winding of the dipole matrix element between the valence and conduction bands composed of gapped chiral fermions (Fig. 2 (g-j)). At higher energies, resonant excitonic effects become more prominent, with considerable deviation between the spectra with and without electron-hole interactions. Intriguingly, as Fig.4 (e-g) shows, excitonic effects above the bandgap are dominated by the repulsive electron-hole exchange interaction. Unlike the direct Coulomb interaction, which scales with the overlap of the electron wavefunctions, the exchange scales with the overlap of the electron and hole states, thus allowing for mixing of electron states of different character, facilitating the intermixing of bulk and surface states.

Next, we turn to the identification of the experimentally observed chiral excitons[27]. Here, we focus on the 3QL $Bi_2Se_3$, which is the thickest slab size that is computationally tractable for our GW-BSE calculations. We need to solve the BSE with at least 18 valence and 14 conduction bands to include all bulk and surface transitions in the energy range of the RSS to SS2 transition. Minima of the surface states occupy only a small part of the Brillouin zone near Γ and require very high k-point sampling to resolve, so we retain only a patch within a radius of 0.028 $a_0^{-1}$ of the Γ point (grey shadow shown in Fig. 1 (c))[55-57]. Solving the BSE Hamiltonian in this smaller patch allows us to increase the k-point sampling to 300×300×1, resolving the surface excitons, which are highly localized in reciprocal space. Since excitons with right and left-circularly polarized selection rules are degenerate, we illustrate their different behavior by calculating the intensity of the *instantaneous* emission (i.e. without considering thermalization) of light with polarization $\sigma'$ following excitation by light of polarization $\sigma$ as $I_{\sigma\sigma'} \propto |<0|\hat{p}_\sigma|S><S|\hat{p}_{\sigma'}|0>|^2$ where $S$ denotes the quantum number of exciton and $\hat{p}_\sigma$ is the momentum operator. We note, however, that in actual photoemission processes, excitons with a large admixture of bulk states will have a shorter non-radiative lifetime and thus be less apparent in photoemission spectra.

Fig. 3 (a) depicts $I_{RL}$ and $I_{RR}$. We identify ten excitonic peaks with circular-polarization-preserving emission, labelled $CX_1$-$CX_{10}$, in the energy window from 0 to 2.82 eV. Of these, $CX_1$, $CX_3$, $CX_8$, and $CX_{10}$ exhibit the greatest circular dichroism and are the focus of the following discussions (see SI for other states). The figure insets show the k-space amplitude of the excitons contributing to each peak, where all identified excitons are well-localized within the patch around Γ. To elucidate the character of these excitons, we break down each exciton state into its component transitions. Fig. 3(b) shows the contributions of each occupied and unoccupied band to a given exciton state, weighted by the degree of circular emission $(I_{RR} - I_{RL})$ [58]. The first circularly polarized emission peak, $CX_1$, which is located at 0.17 eV corresponds to the lowest energy excitons in peak C, discussed previously.

Among the ten identified excitons, $CX_1$ is the only one that is composed entirely of TSS. The other higher energy peaks are resonant states composed of a combination of surface states with chirality-dependent optical transitions and a continuum of lower-energy bulk states. $CX_3$, located at 1.1 eV, primarily arises from transitions between the third highest degenerate set of bulk valence bands (VBM-2) and the surface state SS1' (here, we use prime to identify the upper half of the gapped Dirac cone). Compared to $CX_1$ and $CX_3$, $CX_8$ exhibits reduced circular dichroism. This is because $CX_8$ is composed of 30% RSS-to-(CBM+3) transitions, which have circularly polarized dipole selection rules (see Fig. S3(c)) and 68% non-circularly polarized (VBM-2)-to-(CBM+3) transitions (see Fig. S3(b)). The admixture of the latter significantly relaxes the circular dichroism. The highest-energy chiral exciton peak we identify, $CX_{10}$, which is located at 2.78 eV and mainly consists of the RSS-to-SS2 transitions, preserves a high degree of circular polarization. From the features of $CX_8$ and $CX_{10}$, we can speculate that the experimentally observed broad circularly-polarized photoemission peak around 2.4 eV in previous work [27] may be due to both $CX_8$ and $CX_{10}$.

Finally, to elucidate the effects of the mixing of TSS and non-protected states, we plot the modulus squared of the electronic part of the exciton wavefunction for $CX_3$ and $CX_{10}$, as shown in Fig. 4 (c-d), when the hole is fixed at a position maximizes the electron amplitude. For comparison, two important individual electronic states of SS1' and RSS at the $\Gamma$ point are plotted in Fig. 4 (a-b) (Other electronic and excitonic states may be found in the SI). For SS1', 18% of the electron wavefunction resides on the middle QL. For $CX_1$ (Fig. S5 (c)), which is composed of SS1-to-SS1' transitions, the electron wavefunction on the middle QLs is slightly reduced to 14%, reflecting enhanced localization due to the exciton binding energy compared to the independent-particle picture. For $CX_3$, however, which consists primarily of (VBM-2)-to-SS1' transitions, 32% of the electronic part of the exciton wavefunction extends into the central QL, showing that the excitonic intermixing of the bulk state and surface state results in reduced localization at the surface. In comparison, for $CX_8$ (Fig. S5 (d)), where the electron comes primarily from bulk states, 91% of the electron wavefunction is on the central QL. Finally, Fig. S5 (a) and Fig. 4(b) show that 97% and 92% of the independent particle wavefunctions are localized on the surface QLs for SS2' and RSS. Similarly to $CX_1$, the exciton $CX_{10}$ that mainly comes from the transitions between SS2' and RSS exhibits enhanced localization, with 99% of the electron wavefunction residing on the surface QLs. Thus, electron-hole interactions between two topologically or non-topologically protected surface states can enhance surface localization, while electron-hole interactions between the surface and bulk state reduce the surface localization.

Finally, though our full GW-BSE calculations focus on 3QL $Bi_2Se_3$, the mechanism of exchange-driven intermixing of bulk and surface states is generalizable to $Bi_2Se_3$ of arbitrary thickness and topological surface states in general. Fig. 4(e-g) shows the evolution of the electron-hole exchange matrix elements $\langle vc\boldsymbol{k}|K^x|v'c'\boldsymbol{k}'\rangle$. The matrix elements reveal that the exchange interaction allows for scattering between surface and bulk states at all thicknesses from 3QL to 6QL. The direct Coulomb interaction, which does not allow for significant scattering, is shown in the SI.

In summary, we have performed GW-BSE calculations of the QP bandstructure and optical spectra of bulk, 1QL, 3QL, and 5QL $Bi_2Se_3$, identified multiple series of novel excitonic features arising from transitions between an admixture of topologically protected surface states, topologically trivial surface states and bulk states, and studied their evolution with layer thickness. We characterize the lowest energy bound excitons in 1QL and 3QL $Bi_2Se_3$ and find that the bright excitons are doubly-degenerate with an angular momentum of $m = \pm 1$, consistent with p-like hydrogen atom states. The exciton binding energy decreases with increasing layer number, but exciton effects remain significant in 3QL $Bi_2Se_3$ for resonant states above the QP bandedge, where they are dominated by the electron-hole exchange interaction, which allows for significant mixing of bulk and surface states. We reveal the existence of ten chiral excitons, which arise from surface-state-to-surface-state or surface-state-to-bulk-state transitions, suggesting that thin films of $Bi_2Se_3$ may be an interesting platform for exploring inter-exciton processes involving topological surface states. Furthermore, our calculations suggest that the previously observed broad chiral PL peak can be ascribed to the combined effects of two chiral excitons located at 2.48 and 2.78 eV [27]. These results provide a comprehensive understanding of the chiral excitons of $Bi_2Se_3$, paving the way for engineering chiral excitons for spin optoelectronics.

This work was supported by the U.S. Department of Energy, Office of Science, Basic Energy Sciences under Early Career Award No. DE-SC0021965. D.Y.Q. acknowledges support by a 2021 Packard Fellowship for Science and Engineering from the David and Lucile Packard Foundation. Development of the BerkeleyGW code was supported by Center for Computational Study of Excited-State Phenomena in Energy Materials (C2SEPEM) at the Lawrence Berkeley National Laboratory, funded by the U.S. Department of Energy, Office of Science, Basic Energy Sciences, Materials Sciences and Engineering Division, under Contract No. DE-C02-05CH11231. The calculations used resources of the National Energy Research Scientific Computing (NERSC), a DOE Office of Science User Facility operated under contract no. DE-AC02-05CH11231; the Extreme Science and Engineering Discovery Environment (XSEDE), which is supported by National Science Foundation grant number ACI-1548562; and the Texas Advanced Computing Center (TACC) at The University of Texas at Austin.

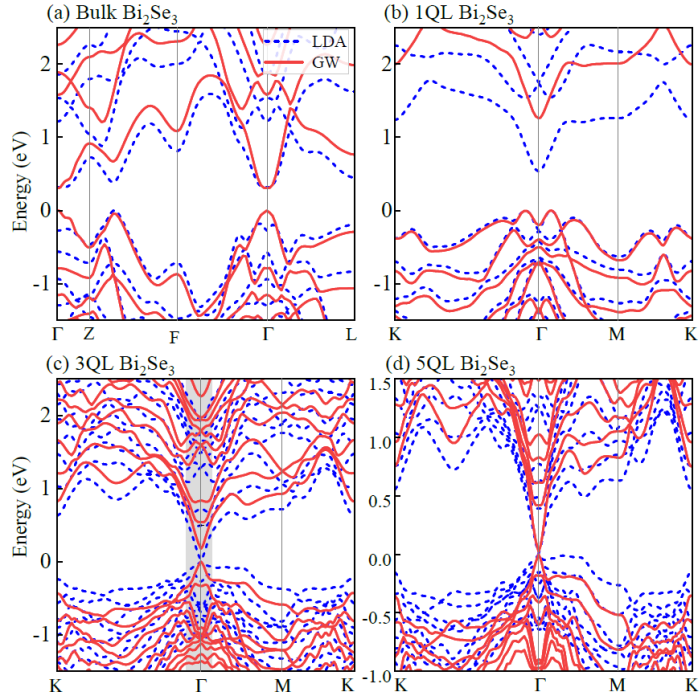

FIG.1 Electronic bandstructure at the $G_0W_0$ (solid red curve) and LDA (dashed blue curve) levels of (a) bulk, (b) 1QL, (c) 3QL, (d) 5QL $Bi_2Se_3$. The grey shadow in (c) shows the extent of a $\Gamma$-centered patch with a radius of $0.028\ a_0^{-1}$.

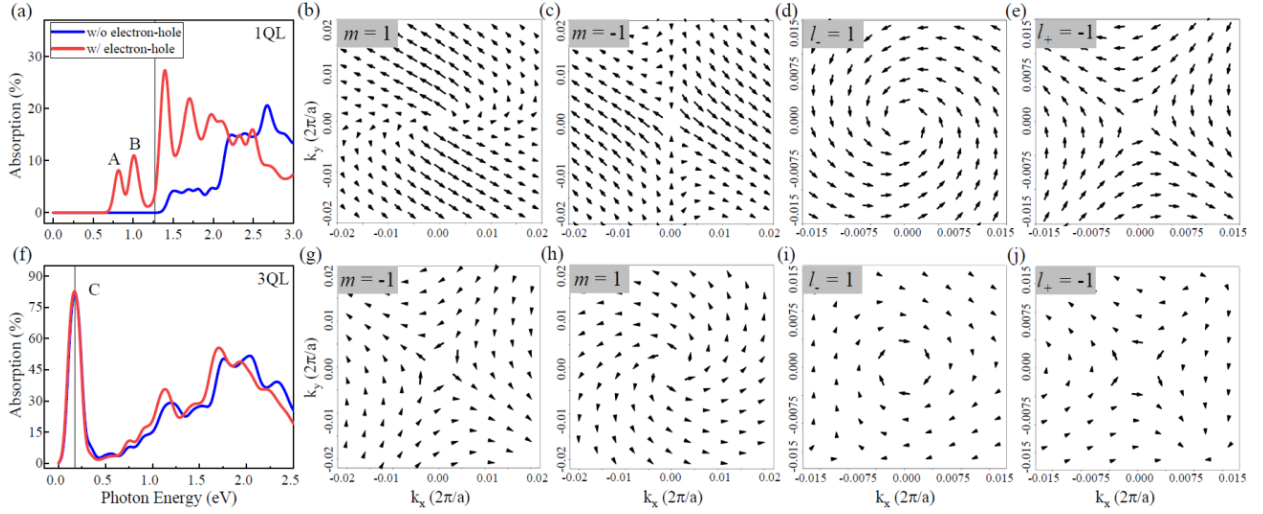

FIG.2 Calculated absorption spectra of (a) 1QL and (f) 3QL $Bi_2Se_3$ with(without) electron-hole interactions included at the GW-BSE(GW-RPA) level in red(blue). The gray vertical line indicates the $G_0W_0$ quasiparticle bandgap. (b-c) The phase winding of the exciton envelope wavefunction, $A^S_{vck}$, of the degenerate bright excitons contributing to peak A of 1QL $Bi_2Se_3$ and (g-h) degenerate bright excitons contributing to peak C of 3QL $Bi_2Se_3$. The angle of the arrows indicates the phase, and the length of each arrow indicates the amplitude. The index $m$ denotes the winding number around the Γ point. (d-e) The winding of the optical interband dipole transition matrix element between the highest valence band and the lowest conduction band for 1QL and (i-j) 3QL $Bi_2Se_3$. The index $l_{-(+)}$ denotes the winding number of left(right) circularly polarized light respectively. A maximally smooth local gauge is imposed following Refs. [59,60]. The exciton wavefunction and optical interband transitions are plotted in a patch centered around the Γ point in the BZ, marked by grey area in Fig. 1(c).

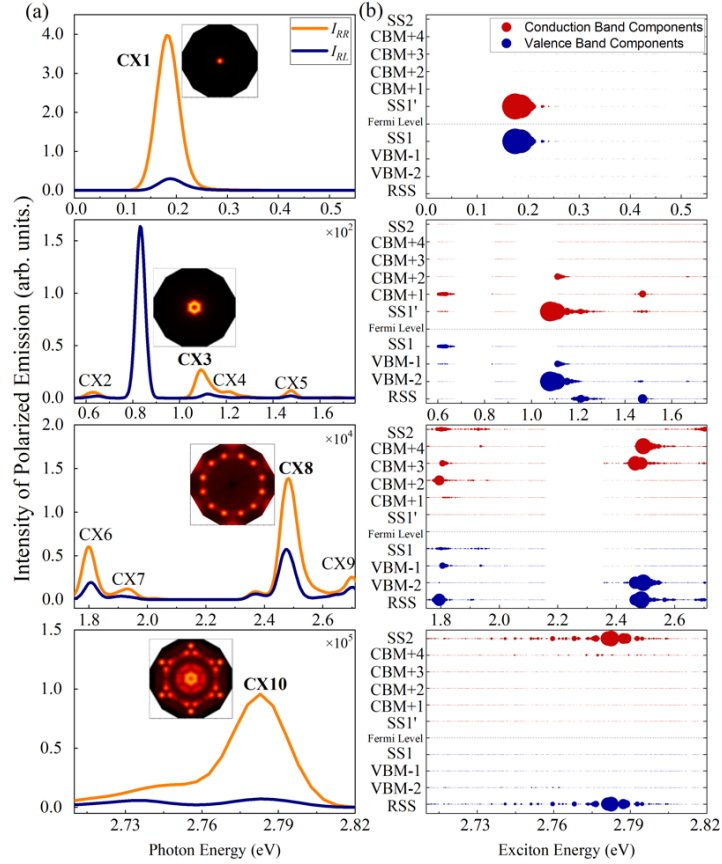

FIG.3 (a) Instantaneous emission intensity of left-hand ($I_{RL}$, dark blue) or right-hand ($I_{RR}$, orange curve) polarized light after excitation by right-hand polarized light from all exciton states in a patch of radius of 0.028 $a_0^{-1}$ around Γ in the BZ. The inset shows the amplitudes of the envelope wavefunction of $CX_1$, $CX_3$, $CX_8$, and $CX_{10}$ in the patch in reciprocal space. (b) Contribution of each band to each exciton state. The bands are labeled as either bulk-like bands relative to the valence band maximum (VBM) and conduction band minimum (CBM) or as surface states SS1, SS2, and RSS, with the occupied band contributions in blue and unoccupied band contributions in red. Each bulk band is doubly degenerate, while surface states sum contributions from all the occupied or unoccupied surface states. The size of each dot is proportional to $(I_{RR} - I_{RL}) \times \sum_{ck} |A_{vck}^S|^2$ for occupied states and $(I_{RR} - I_{RL}) \times \sum_{vk} |A_{vck}^S|^2$ for unoccupied states.

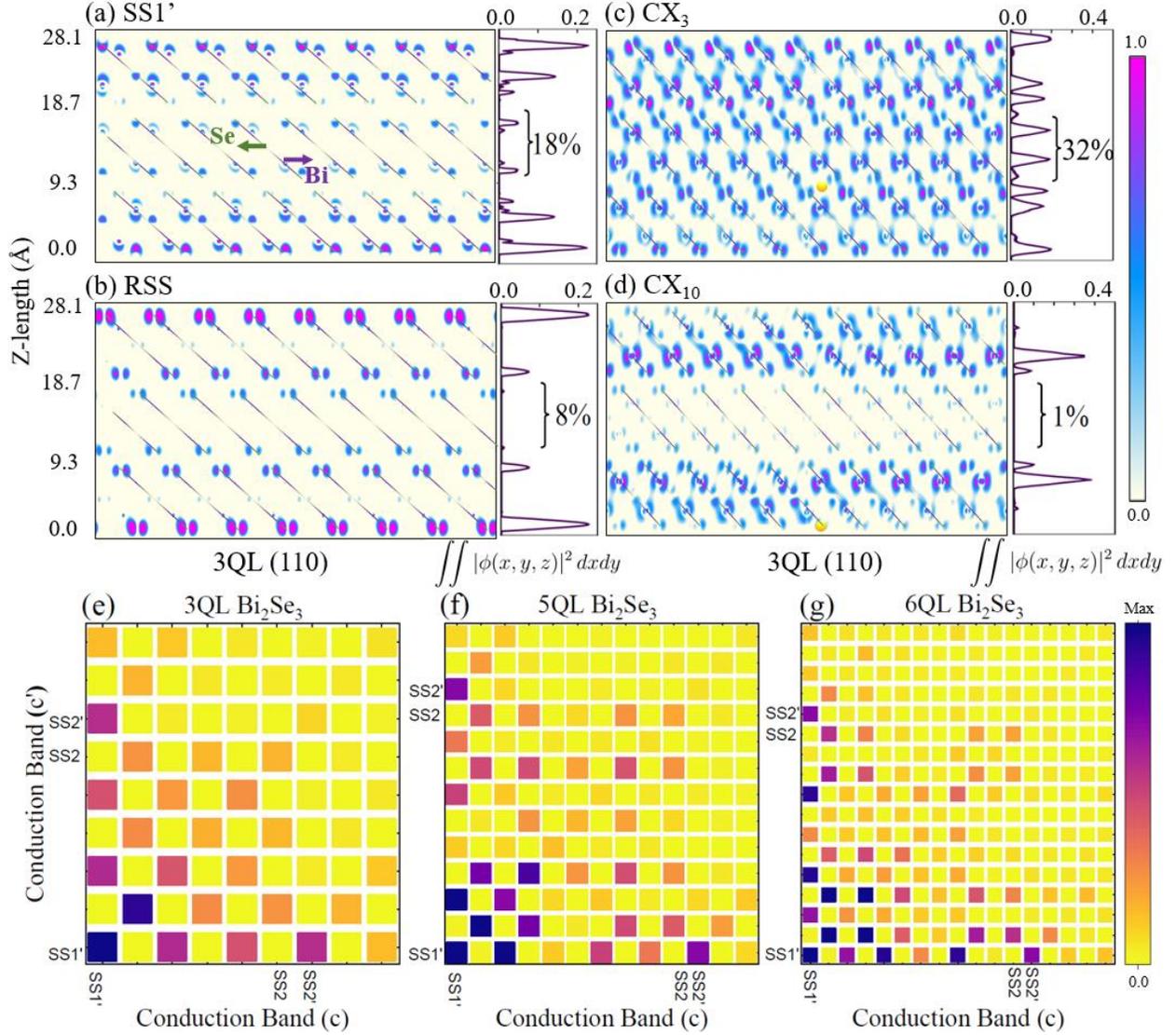

FIG.4 (a-b) The modulus of the electronic wavefunction in 3QL Bi$_2$Se$_3$ for the surface states SS1' and RSS at the Γ point. (c-d) The modulus of the exciton wavefunction for CX$_3$ and CX$_{10}$, when the hole (yellow sphere) is fixed. The hole position is set at the location that maximizes the wavefunction amplitude. The integral of the modulus squared of each wavefunction over the *x-y* plane is shown as a side panel in (a-d). The percentage label indicates the contribution to the wavefunction from the central QL. For clarity, the Bi and Se atoms are not explicitly shown. The bonds are depicted as purple lines in the vicinity of Bi and green lines in the vicinity of Se. (e-g) Electron-hole exchange matrix elements $\langle vc\boldsymbol{k}|K^x|v'c'\boldsymbol{k}'\rangle$ for $\boldsymbol{k} = \boldsymbol{k}' = \Gamma$ and $v = v' = $ SS1 for (e) 3QL, (f) 5QL and (g) 6QL slabs of Bi$_2$Se$_3$.